\documentclass[fdp,fleqn]{w-art}
\usepackage{times}
\usepackage{w-thm}

\usepackage{latexsym}
\usepackage{epsfig,amssymb,euscript}
\usepackage{amsmath,cancel,slashed}
\def\d {\text{d}}
\def\tr {\text{tr}}
\newcommand{\ul}{\underline}

\begin{document}
\pagespan{3}{}



\title[Heterotic DWSB]{DWSB for heterotic flux compactifications}


\author[J. Held]{Johannes Held\inst{1,}%
  \footnote{E-mail:~\textsf{heldj@mppmu.mpg.de}, 
            Phone: +00\,49\,89\,32354-405, 
            Fax: +00\,49\,89\,3226704}}
\address[\inst{1}]{Max- Planck- Institut f\"ur Physik, F\"ohringer Ring 6, 80805 M\"unchen, Germany}
\begin{abstract}
We investigate the construction of non-supersymmetric vacua in compactifications of heterotic string theory with intrinsic torsion and background fluxes. We do this by using the technique of \textit{domain-wall supersymmetry breaking} (DWSB) that was developed in the context of type II compactifications. By means of a scalar potential we derive conditions on the compactification manifold that must hold in order to satisfy the equations of motion up to order $\alpha'$.
\end{abstract}
\maketitle                   





\section{Introduction}
During the last decade a lot of effort has been put into the understanding of string theory compactifications including background fluxes. Especially supersymmetric settings are up to now well understood (see \cite{Grana:2005jc,Douglas:2006es,Blumenhagen:2006ci} and references therein).

Nonetheless, in order to make contact to particle physics an important issue is how to incorporate supersymmetry breaking into string theory. For the heterotic string a popular mechanism is to achieve the breaking by gaugino condensation on a hidden gauge group  \cite{din85,drsw85}. In the setting of type II string theory there is also the possibility to induce SUSY breaking by the inclusion of background fluxes \cite{gkp}. In \cite{dwsb} clear conditions where given which a non-supersymmetric type II vacuum has to satisfy to be consistent. However, also in the heterotic case one can turn on non-trivial background fluxes and hence by duality arguments find consistent non-supersymmetric vacua \cite{Becker:2009zx}. This leaves open the question if these vacua can be constructed directly from the ten-dimensional heterotic supergravity action, which we answered affirmatively in \cite{Held:2010az}, using the techniques of \cite{dwsb}. This article is a short review of this work.

\section{A BPS- like potential}
\subsection{The ten-dimensional action}
The bosonic action up to order $\mathcal{O}(\alpha')$ of ten-dimensional $\mathcal{N}=1$ heterotic supergravity is given by \cite{bdr}
\begin{equation}
\label{10daction}
S=\frac{1}{2\kappa^2}\int\d^{10} x\sqrt{-{\text{ det}\, } g}\,e^{-2\phi}\big[\mathcal{R}_{X_{10}}+4(\d\phi)^2-\frac{1}{2} H^2+\frac{\alpha^\prime}{4}(\tr R_+^2- \tr F^2)\big]~.
\end{equation}
here $\mathcal{R}_{X_{10}}$ is the scalar curvature of the ten-dimensional space, $\phi$ the dilaton, $F$ the gauge field strength and $R_+$ the curvature built from the torsionful connection\footnote{All conventions used are exactly as in \cite{Held:2010az} and can be found there.}
\begin{equation}
\label{plusconnection}
\omega^{\ul{M}}_\pm{}_{\ul{N} P}=\omega^{\ul{M}}{}_{\ul{N} P}\pm\frac12 H^{\ul{M}}{}_{\ul{N} P}~.
\end{equation}
The Neveu--Schwarz (NS) three-form flux $H$ satisfies then the Bianchi identity (BI) with respect to $R_+$
\begin{equation}
\label{BI}
\d H=\frac{\alpha^\prime}{4} (\tr R_+\wedge R_+-\tr F\wedge F)~.
\end{equation}
Combining the equations of motion for the metric and the dilaton one finds the 'modified' Einstein equation
\begin{equation}
\label{modEinst}
R_{PQ}+2\nabla_P\nabla_Q\phi-\frac12\iota_P H\cdot\iota_Q H +\frac{\alpha^\prime}{4}\big[\tr(\iota_P R_+\cdot \iota_Q R_+)  -\tr (\iota_P F\cdot\iota_Q F)\big]=0
\end{equation}
\subsection{Compactification ansatz}
\label{CompAnsatz}
In order to make contact to four-dimensional physics we choose our ten-dimensional space to be of the form $X_{10}=X_{4}\times_w M$. Here, $X_4$ denotes a maximally symmetric space with cosmological constant $\Lambda$, and $M$ a compact six-dimensional manifold equipped with an SU$(3)$-structure \cite{Chiossi:2002tw}, respectively. The ten-dimensional metric takes then the form
\begin{equation}
\d s^2_{X_{10}}=e^{2A}\d s^2_{X_{4}}+\d s^2_{M}~,
\end{equation}
where the warp factor $A$, as well as the fields $\phi$, $H$, and $F$, depend only on the internal space $M$. This in turn implies that in the BI (\ref{BI}) $R_+$ also reduces to its equivalent on $M$.

This ansatz alone has already far reaching consequences. Restricting the indices $P$ and $Q$ in (\ref{modEinst}) to $X_4$ one finds the equation
\begin{eqnarray}
\label{extEinstcompl}
\nabla^m(e^{-2\phi}\nabla_m e^{4A})&=& 4 \, e^{2A-2\phi}\Lambda+\alpha^\prime\, e^{2A-2\phi}\Big\{\frac{2}3\, e^{-2A}[\Lambda-3(\d A)^2]^2\cr && +2(\nabla_m\nabla_n e^A)(\nabla^m\nabla^n e^A)+ (\iota_m H\cdot \iota_n H)\, \nabla^m e^A\nabla^n e^A\Big\}~,
\end{eqnarray}
which can be solved iteratively in $\mathcal{O}(\alpha')$ and yields that up to $\mathcal{O}(\alpha'\,^3)$ the cosmological constant vanishes and the warp factor is constant. Since our analysis is valid only up to order $\alpha'$ we are allowed to consider $X_4$ as a Minkowski space with constant warping.
\subsection{Supersymmetry and BPS-like potential}
\label{SUSYsection}
To analyze the consequences of our compactification ansatz more deeply we will follow the approach of \cite{dwsb} and consider an effective four-dimensional potential, from which the equations of motions for all fields can be derived. Using the fact that d$A$ and $\Lambda$ have to be zero in our setting we can write the action as $S=-\int_{X_4}d^4x V$ with
\begin{eqnarray}
\label{firstpot}
V&=&\frac{1}{2\kappa^2}\int_M \text{vol}_M\ e^{4A-2\phi}\Big[-\mathcal{R}+\frac{1}{2} H^2-4(\d\phi)^2+\frac{\alpha^\prime}{4} (\tr F^2-\tr R^2_+)\Big]~,
\end{eqnarray}
where now $\mathcal{R}$ denotes the Ricci scalar of the internal manifold $M$. The equation of motion of $A$ implies then that $V=0$ on-shell, which is also demanded by the fact that we consider four dimensional Minkowski vacua.

In order to understand how supersymmetry breaking can be analyzed starting from this potential we make use of the fact that $M$ is an SU$(3)$-structure manifold. In particular, as was shown in \cite{Gauntlett:2003cy}, given the SU$(3)$-structure forms $J$ and $\Omega$ the conditions for supersymmetry read
\begin{subequations}
\label{susycond}
\begin{align}
&\d(e^{-2\phi}\Omega)=0 \label{susycond1}\\
&\d(e^{-2\phi}J\wedge J)=0 \label{susycond3}\\
&e^{2\phi}\d(e^{-2\phi}J)=*H~. \label{susycond2}
\end{align}
\end{subequations}
These conditions can also be understood as calibration conditions on NS5-branes \cite{Martucci:2005ht} wrapping three-, four-, and two-cycles in $M$ respectively, and are hence perceived as domain walls, strings, and space-time filling gauge theories in four dimensions.

Since $J$ and $\Omega$ contain all information encoded in the metric of $M$ it is possible to rewrite the scalar curvature $\mathcal{R}$ in terms of these forms \cite{dwsb,bedulli06,Cassani:2008rb}
\begin{eqnarray}
\label{curvSU(3)}
{\cal R}=-\frac{1}{2}(\d J)^2-\frac18[\d(J\wedge J)]^2-\frac{1}{2}|\d \Omega|^2+\frac{1}{2}|J\wedge \d\Omega|^2+\frac{1}{2} u^2-\nabla^mu_m
\end{eqnarray}
with the one-form $u$ given by
\begin{equation}
u=u_m\d y^m=\frac{1}{4}(J\wedge J)\lrcorner\d(J\wedge J)-\frac{1}{2}\text{Re}(\bar\Omega \lrcorner \d\Omega)~.
\label{curvSU(3)b}
\end{equation}
Using this formula and the BI (\ref{BI}) one can rearrange the potential (\ref{firstpot}) such that it is written only in terms of squares. Separating $\mathcal{O}(\alpha'^0)$ from $\mathcal{O}(\alpha')$ terms one gets $V=V_0+V_1$ with
\begin{subequations}
\label{pot2}
\begin{align}
V_0=&\frac{1}{4\kappa^2}\int \text{vol}_M\ e^{4A-2\phi}\big[e^{2\phi}\d(e^{-2\phi} J)-*H\big]^2\,+\,\frac{1}{4}\big[e^{2\phi}\d\big(e^{-2\phi}J\wedge J \big) \big]^2\label{potord0}\\
+ & \frac{1}{4\kappa^2}\int \text{vol}_M\,e^{4A-2\phi}\Big[|e^{2\phi}\d(e^{-2\phi}\Omega)|^2-|J\wedge e^{2\phi}\d(e^{-2\phi}\Omega)|^2\Big]\nonumber\\
-& \frac{1}{4\kappa^2}\int \text{vol}_M\,e^{4A-2\phi}\Big\{\frac1{4}e^{2\phi}(J\wedge J)\lrcorner \d(e^{-2\phi}J\wedge J)\,+\,\frac{1}{2}e^{2\phi}\,\text{Re}[\bar\Omega\lrcorner \d(e^{-2\phi}\Omega)]\Big\}^2\nonumber\\
V_1=& \frac{\alpha^\prime}{8\kappa^2}\int \text{vol}_M e^{4A-2\phi}\big[2\text{Tr}\, |F^{2,0}|^2+\text{Tr}\,|J\cdot F^{1,1}|^2\big] -\big[2\text{Tr}\, |R_+^{2,0}|^2+\text{Tr}\,|J\cdot R^{1,1}_+|^2\big]~.
 \label{potord1}
\end{align}
\end{subequations}
Note that all terms appearing at zeroth order in $\alpha'$ (\ref{potord0}) will vanish quadratically when the supersymmetry conditions (\ref{susycond}) are implied. On the other hand the terms appearing in $V_1$ vanish quadratically if $F$ as well as $R_+$ are primitive $(1,1)$-forms, i.e.
\begin{equation}
\label{primitivity}
F^{2,0}~=~R_+^{2,0}~=~J\lrcorner F~=~J\lrcorner R_+~=~0~.
\end{equation}
But, as was shown in \cite{Sen:1986mg}, these equations are also implied by supersymmetry. Thus we find that the effective potential of heterotic supergravity compactified on an SU$(3)$-structure manifold can be put in a BPS-like form, by which we mean that it and its variations vanish once supersymmetry is imposed. Differently speaking, we have shown that indeed supersymmetry and the BI are sufficient to satisfy the equations of motion. The next question to address is how this picture changes once supersymmetry is broken.
\section{Heterotic domain-wall SUSY breaking}
\subsection{Conditions}
In order to keep the effects of supersymmetry breaking tractable as well as controllable we will keep as much of the conditions of section \ref{SUSYsection} as possible. In fact we will allow that
\begin{equation}
\label{DWSB}
\d(e^{-2\phi}\Omega)~\neq~0~,
\end{equation}
while we still demand the following conditions
\begin{subequations}
\label{DWSBcond}
\begin{align}
&\d(e^{-2\phi}J\wedge J)=0\\
&e^{2\phi}\d(e^{-2\phi}J)=\ast H\\
&\bar{\Omega}\lrcorner\d (e^{-2\phi}\Omega)=0~.
\end{align}
\end{subequations}
This means that NS5-branes wrapping three-cycles of $M$ and hence appearing as domain-walls in $X_4$ are not any longer calibrated. Thus this type of supersymmetry breaking was named \textit{domain-wall SUSY breaking} in \cite{dwsb}. In more geometric terms allowing (\ref{DWSB}) means that $M$ is not any longer a complex manifold but only almost complex with respect to $J$ and $\Omega$.
\subsection{Implications}
To analyze the implications of this ansatz we will focus first on $V_0$, the $\mathcal{O}(\alpha'^0)$ part of the potential (\ref{pot2}). First one should notice that the reasoning of section \ref{CompAnsatz} does not depend on any supersymmetry arguments. Hence in particular $A$ is still constant and the potential should vanish on-shell. Imposing the conditions (\ref{DWSBcond}) we arrive at
\begin{equation}
V_0^\prime~=~\frac{1}{4\kappa^2}\int \text{vol}_M\,e^{4A-2\phi}\big[|e^{2\phi}\d(e^{-2\phi}\Omega)|^2-|J\wedge \d \Omega|^2\big]~\stackrel{!}{=}~0~,
\label{potpiece}
\end{equation}
which is equivalent to demanding
\begin{equation}
\label{sbcond}
|e^{2\phi}\d(e^{-2\phi}\Omega)|^2=|J\wedge \d\Omega|^2~.
\end{equation}
Expanding d$J$ and d$\Omega$ in terms of torsion classes \cite{Chiossi:2002tw} we find that (\ref{DWSBcond}) implies
\begin{equation}
e^{2\phi}\d(e^{-2\phi}\Omega)=W_1\, J\wedge J+W_2\wedge J
\end{equation}
and that the condition (\ref{sbcond}) is satisfied if 
\begin{equation}
\label{finalcond1}
|W_2|^2~=~24\,|W_1|^2
\end{equation}

Since the two terms of $V_0'$ in (\ref{potpiece}) do not vanish separately one also has to make sure that the equations of motion coming from this potential are indeed satisfied. Defining a two form $\mathcal{S}:=W_2+4\,W_1\,J$ we find as an additional constraint
\begin{eqnarray}
\label{eomtor}
\text{Im}\big[\iota_{(m}\bar\Omega\cdot\iota_{n)}\d\mathcal{S}\big]&=&
8g_{mn}|W_1|^2-2\text{Re}[\bar W_1(\iota_m W_2\cdot \iota_n J)]-\text{Re}[\iota_{m}W_2\,\cdot\,\iota_{n}\bar{W}_2]\\
&=& |W_1|^2\Big\{9g_{mn}-\text{Re}\Big[ \iota_m\Big(\frac{W_2}{W_1}+J\Big)\cdot\iota_n\Big(\frac{\bar W_2}{\bar W_1}+J\Big) \Big]\Big\}\nonumber~.
\end{eqnarray}
One should note that our SUSY breaking mechanism is governed by one scalar parameter, namely $W_1$. Since $W_1$ has mass dimension 1 we relate it to the supersymmetry breaking scale $W_1\sim M_{SB}$ and SUSY breaking effects remain small as long as $W_1$ is kept sufficiently small.

Turning to $V_1$ one finds that the conditions for its vanishing are the same as in the supersymmetric case (\ref{primitivity}). But only the primitivity of $F$ is still guaranteed, while $R_+$ could in principle be $(2,0)$ as well as non-primitive. Nevertheless, for our analysis it suffices that $R_+^{2,0}$ and $J\lrcorner R_+$ vanish at zeroth order in $\alpha'$, since they appear only in the first order correction to the potential which is the highest order of correction that we study.

A detailed analysis of the supersymmetry variation of the gravitino, as can be found in \cite{Held:2010az}, shows then that $R_+$ is only then $(1,1)$ and primitive to a sufficient order in $\alpha'$ if the SUSY breaking scale $M_{SB}$ is much smaller then the compactification scale $M_{KK}$
\begin{equation}
M_{SB}~\ll~M_{KK}~.
\end{equation}

To conclude, in order to have a consistent DWSB setting there are several condition to be met. First one has to make sure that the conditions (\ref{finalcond1}) and (\ref{eomtor}) coming from the zeroth order part of the potential are satisfied. This should be possible by choosing a  suitable manifold $M$. Then one has to make sure that the SUSY braking scale defined by $W_1$ is much smaller then the compactification scale while one still has to satisfy the Bianchi identity (\ref{BI}).

As is shown in more detail in \cite{Held:2010az} these conditions can indeed be satisfied for elliptic fibrations of a warped K3 manifold, also analyzed in \cite{Becker:2009zx,Dasgupta:1999ss,Fu:2006vj}. The need to fulfill the Bianchi identity and the quantization of the $H$-flux leads then to a stabilization of the dilaton $\phi$ and the volume of the elliptic fiber. Moreover, it can be shown that the gravitino mass is inverse proportional to the volume of K3 and hence that supersymmetry breaking is weak for K3 sufficiently large.

\begin{acknowledgement}
I would like to thank the organizers of the XVIth European Workshop on String Theory for the opportunity to present my work there. I also would like to thank Dieter L\"ust, Fernando Marchesano, and Luca Martucci for the fruitful collaboration.
\end{acknowledgement}

\end{document}